\documentclass[preprint,12pt]{elsarticle}

\usepackage{amssymb}
\usepackage{tabularx}
\usepackage{longtable}
\usepackage{multirow}
\usepackage{multirow}
\usepackage{pgfplots}
\usepackage{tikz}
\usepackage{lscape}
\usepackage{amsmath}

\usepackage{arydshln}  % dashed lines

\usepackage{caption} % Required for \captionsetup
\usepackage{booktabs} % For better table formatting
\usepackage[colorlinks=true,allcolors=blue]{hyperref} % Generate hyper-links to the citations

\newcolumntype{Y}{>{\centering\arraybackslash}X}
\newcolumntype{P}[1]{>{\centering\arraybackslash}p{#1}}
\newcolumntype{M}[1]{>{\centering\arraybackslash}m{#1}}
\usepackage{tikz}
 
\usepackage{pifont}
\newcommand{\cmark}{\ding{51}}%

\pgfplotsset{compat=1.18} 
\usepackage{comment}

\newif\ifappendixLandscape
\usepackage{listings}
\usepackage{xcolor}

\journal{}

\begin{document}

% ############################################
%           VARIABLES 
% ############################################
%\def \totalworks{121}
\def \inscopeworks{ 62 }

\def \secanalysisworks{4} % OK
\def \empiricalworks{7} % OK
\def \usecaseworks{21}  % OK 
\def \attackscenarioworks{9} % OK 
\def \enhancementworks{27} % OK 
\def \attackdetectionworks{12} % OK 
\def \vulndiscworks{16} % OK 
\def \otherworks{25} % OK 

\def \totalworks{\the\numexpr\secanalysisworks + \empiricalworks + \usecaseworks + \attackscenarioworks + \enhancementworks + \attackdetectionworks + \vulndiscworks + \otherworks \relax}

\def \categorizedworks{\the\numexpr \totalworks - \otherworks \relax}

\begin{frontmatter}

\title{WebAssembly and Security: a review}

\author[inst1]{Gaetano Perrone}
\author[inst1]{Simon Pietro Romano}

\affiliation[inst1]{organization={University of Naples Federico II, Department of Electrical Engineering and Information Technology},
            addressline={Via Claudio 21}, 
            city={Naples},
            postcode={80125}, 
            state={Naples},
            country={Italy}}

\begin{abstract}
WebAssembly is revolutionizing the approach to developing modern applications. Although this technology was born to create portable and performant modules in web browsers, currently, its capabilities are extensively exploited in multiple and heterogeneous use-case scenarios. 
With the extensive effort of the community, new toolkits make the use of this technology more suitable for real-world applications.   
In this context, it is crucial to study the liaisons between the WebAssembly ecosystem and software security.
Indeed, WebAssembly can be a medium for improving the security of a system, but it can also be exploited to evade detection systems or for performing crypto-mining activities. In addition, programs developed in low-level languages such as C can be compiled in WebAssembly binaries, and it is interesting to evaluate the security impacts of executing programs vulnerable to attacks against memory in the WebAssembly sandboxed environment. Also, WebAssembly has been designed to provide a secure and isolated environment, but such capabilities should be assessed in order to analyze their weaknesses and propose new mechanisms for addressing them. 
Although some research works have provided surveys of the most relevant solutions aimed at discovering WebAssembly vulnerabilities or detecting attacks, at the time of writing, there is no comprehensive review of security-related literature in the WebAssembly ecosystem.
We aim to fill this gap by proposing a comprehensive review of research works dealing with security in WebAssembly. We analyze~\totalworks~papers by identifying seven different security categories.

We hope that our work will provide insights into the complex landscape of WebAssembly and guide researchers, developers, and security professionals towards novel avenues in the realm of the WebAssembly ecosystem.
\end{abstract}


\end{frontmatter}


\section{Introduction}
WebAssembly is a formal specification for portable machine code, born to allow the realization of portable code developed in any language and executed by modern browsers. 
This technology was developed to provide a client-side solution for browsers, but several cloud providers are nowadays offering WebAssembly runtime environments for the server side. This reveals the key innovations that WebAssembly promotes in terms of performance, isolation, distribution, and modularity. 
With the spread diffusion of WebAssembly, it becomes crucial to research the liaison between this technology and security. Security research can involve different topics, such as empirical studies for evaluating the presence of security flaws in WebAssembly design and systems, the realization of solutions aimed at enhancing WebAssembly security, the implementation of novel approaches for discovering vulnerabilities or detecting attacks in WebAssembly, or the proposal of WebAssembly solutions for security-related use cases.
Although several studies have been developed in the above-mentioned research fields, to the best of our knowledge, there is no work that provides an exhaustive survey of the existing literature.

This work aims to contribute to this direction by providing a comprehensive review of the most relevant works that have investigated the liaisons between security and the WebAssembly ecosystem.  We analyze~\totalworks~works, classifying~\categorizedworks~ into seven security categories and describing the remaining~\otherworks~additional works that, while not fitting into a specific security category, provide fundamental knowledge for expanding the literature on WebAssembly and security. We hence provide a security overview of the WebAssembly world, which will hopefully represent a starting point for researchers looking to investigate future works, as well as for security practitioners aiming to integrate this appealing technology into their systems.

The remainder of this article is structured in six sections.
Section~\ref{sec:preliminary} conveys to the readers the preliminary concepts functional to follow subsequent sections.
Section~\ref{sec:search} illustrates the review process we followed for retrieving works relevant to our review.
Section~\ref{sec:related} shows existing studies that reviewed WebAssembly research works, with particular emphasis on those that have concentrated their efforts on security, by illustrating the differences with our findings. 
Section~\ref{sec:security-review} describes the reviewed works, while Section~\ref{sec:discussion} provides a discussion on these by summarizing relevant information and research gaps that could be addressed in future investigations. Section~\ref{sec:conclusions} concludes our research path by proposing insights for researchers who want to explore challenging avenues in the  WebAssembly security field.
\section{Preliminary concepts}
\label{sec:preliminary}
In this section, we first describe the WebAssembly ecosystem by introducing the reader to the basics. Then, we introduce the concepts that can be useful to understand the relevant works that will be illustrated in Section~\ref{sec:security-review}, namely, smart contracts and security techniques for addressing low-level vulnerabilities.

\subsection{WebAssembly Introduction}
In 2015, Brendan Eich announced the creation of WebAssembly, an innovative approach to writing assembly programs for the web \cite{eich-wasm} aimed to replace \texttt{asm.js} \cite{asmjs}.
The main reasons behind the realization of an alternative approach to \texttt{asm.js} were to simplify the compiler options and to increase performance by natively decoding code much faster than JavaScript  \cite{wasm-faq}.
In 2017, the WebAssembly Community Group was created with the mission of providing collaboration on defining a pre-standardization of WebAssembly, and in 2019, WebAssembly became a World Wide Web Consortium (W3C) recommendation \cite{wasm-core}.
% DESIGN CONCEPTS 
Design goals are focused on: 
\begin{itemize}
    \item \textbf{Performance}: the code execution should be nearest to native code performance by taking advantage of modern underlying hardware features; 
    \item \textbf{Security}: the code should be validated and executed in memory-safe, isolated, and sandboxed environments; 
    \item \textbf{Portability}: the code should be independent from language, hardware, and platform.
\end{itemize}

After the WebAssembly foundation, several companies and nonprofit organizations such as Bytecode Alliance~\cite{bytecode} were interested in supporting the project.

% HETEROGENEOUS LANGUAGES
WebAssembly was designed to build C/C++ projects, but today, there is a strong community effort to increase the number of high-level languages supported.
At the time of writing, it is possible to compile Python, Ruby, PHP, C, and Rust programs and generate WebAssembly code that can be executed into a WebAssembly compiler.

As regards the ecosystem, some companies are following the approach of the Docker community, which developed DockerHub, an open search engine and repository for finding ready-to-use Docker images \cite{9240654}. In particular, the Wasmer team provides a complete WebAssembly framework, constituted by a WebAssembly runtime for running WebAssembly binaries~\cite{wasmer}, a Software Development Kit (SDK) to embed WebAssembly programs in any programming language~\cite{wasmer-sdk}, a serverless platform for executing programs developed through the SDK~\cite{wasmer-edge}, and a publicly available and searchable repository where it is possible to find any WebAssembly binary developed by the community~\cite{wasmer-hub}. 

\subsubsection{WebAssembly components}
The WebAssembly ecosystem consists of several components that enable the execution of programs (at runtime or compiled) written in multiple languages. WebAssembly used to be mainly focused on executing C programs, but currently, it supports the entire set of most popular programming languages. The developers write a program in their own preferred language. Then, the source code can follow two processing steps: (i) it can be compiled by a \textit{WebAssembly compiler} and executed by a runtime engine, or (ii) it can be interpreted, i.e., a \textit{WebAssembly interpreter} at runtime processes the lines of code and executes them. The execution is orchestrated by a \textit{WebAssembly runtime engine}, that abstracts the execution from the underlying system (i.e., the browser, the Operating System, etc.) and finally executes either the WebAssembly binary file (in case of compilation) or the current line of code (in case of interpretation) on the underlying system architecture.

\begin{comment}

Figure~\ref{fig:wasm-components} depicts the primary WebAssembly components.
%WASI
    \begin{figure}[ht!]
    \centering
    \includegraphics[width=0.70\textwidth]{images-wasm.png}

    \caption{WebAssembly components}
    \label{fig:wasm-components}
    \end{figure}

\end{comment}

The development of WebAssembly was motivated by the need to provide a low-level assembly-like language for compiling arbitrary source code and running it in the browser. As WebAssembly shows a high level of portability, it has been extended to include other platforms. The WebAssembly System Interface (WASI) \cite{wasi} comprises a collection of  standardized API specifications aimed at providing secure interfaces for programs, thereby extending the capabilities of WebAssembly. The API facilitates interaction with underlying functionality such as I/O, clocks, filesystem, HTTP driver, and more.

% WebAssembly module
WebAssembly programs are organized into modules, i.e., units of deployment, loading, and compilation, that have definitions for types, functions, tables, memories, and globals. Each module declares imports and exports, and upon instantiation, a \textit{start function} is automatically invoked.

\subsubsection{WebAssembly memory layout and binary format}
% https://wasmbyexample.dev/examples/webassembly-linear-memory/webassembly-linear-memory.rust.en-us.html
% 

WebAssembly follows the principle of simplicity by defining a simple linear memory structure, i.e., a mutable array of raw bytes. Each memory has a \emph{metadata region} that keeps information about the memory layout, size, and properties of the memory in the WebAssembly module. Each program loads and stores values from/to a linear memory at any byte address.

The computational model of WebAssembly is based on a \emph{stack machine}. The program's instructions are executed in order and manipulate values by popping arguments and pushing results into the stack. Each instruction is encoded as \textit{opcodes} of one byte (multiple bytes for control instructions), followed by immediate arguments if present. There are several instruction types, namely, control, reference, parametric, variable, table, memory, numeric, vector, and expression.

WebAssembly has two concrete representations, which map to a common structure: 

\begin{itemize}
    \item \textbf{text format ($.wat$ extension)}: designed to let developers understand the structure of a WebAssembly module. It can be used for writing code and compiling it in binary format;
    \item \textbf{binary format ($.wasm$ extension)}: a compact binary instruction format for a stack-based virtual machine. It is the compilation target of higher-level languages.    
\end{itemize}

% Program layout https://webassembly.github.io/spec/core/syntax/modules.html

\subsubsection{WebAssembly use-cases}
\label{sec:wasm-usecases-non-security}
WebAssembly is designed to ensure high flexibility and portability. As an example, a program written in C can be compiled and executed in any architecture, in the browser, on x86 desktop machines, and in arm-embedded microcontrollers. Typical use cases are focused on browser-based applications such as games, peer-to-peer applications, or image recognition, but there are also other application scenarios~\cite{wasm-usecases}. Hilbig et al. (2021)~\cite{10.1145/3442381.3450138} sample 100 random WebAssembly binaries used on the web and underline that WebAssembly is used for gaming, text, and media processing, visualization/animation, as well as demonstrations of programming languages. A relevant report is provided by the Cloud Native Computing Foundation, that every year summarizes the state of WebAssembly in the world~\cite{cncf}. According to their study:
\begin{itemize}
    \item The principal use-cases of WebAssembly are web applications (58\%), data visualization (35\%), Internet of things, games, and backend services; 
    \item Developers are interested in performance benefits, exploring new use cases and technology, and sharing code between projects. They confirm performance benefits when WebAssembly is adopted;
    \item WebAssembly is used both for realizing new applications and migrating existing applications;
    \item The source code languages used for compiling WebAssembly binaries are primarily JavaScript, C\#, C++, and Python.     
\end{itemize}

In our work, we highlight the most pertinent cybersecurity-focused use cases documented in the literature (see Section~\ref{sec:use-cases}).

\subsection{Low-level vulnerabilities in software}
WebAssembly is an assembly language typically used to compile basic code developed in low-level languages such as C. Therefore, it is crucial to introduce foundational concepts surrounding low-level vulnerabilities that could impact a program. In order to understand this kind of vulnerability, it is essential to offer an overview of program execution in machines.

Low-level languages such as C and C++ can lead to various vulnerabilities stemming from unsafe memory management practices:
\begin{itemize}
    \item \emph{Stack-based overflow}: this vulnerability occurs when an attacker can overwrite a memory buffer located in the stack, including function return addresses, leading to arbitrary code execution;
    \item \emph{Heap-based overflow}: this vulnerability is similar to stack overflow, but it occurs in the heap memory section;
    \item \emph{Integer overflow}: this vulnerability occurs when the result of an arithmetic operation exceeds the memory size allocated for the variable holding it, potentially leading to memory corruption or unintended code execution.
\end{itemize}

These vulnerabilities often occur when unsafe functions are input-controllable, allowing a malicious user to send input that overwrites memory buffers.

\subsubsection{Security approaches to discover software vulnerabilities}

Software vulnerability discovery techniques can be classified into two broad categories: 
\begin{itemize}
    \item \emph{Static analysis}: involves source code analysis without executing the program. It is useful to detect vulnerabilities caused by improper input validation or type mismatches, but cannot discover runtime errors or vulnerabilities caused by complex interactions with the code; 
    \item \emph{Dynamic analysis}: these approaches try to find vulnerabilities in programs at runtime by injecting malicious inputs (fuzzing) or by analyzing the code during execution (tracing or symbolic execution). 
\end{itemize}

These approaches can be further classified into several techniques, as illustrated in Table~\ref{tab:security-approaches}.

% LONGTABLE
\bgroup
\renewcommand{\arraystretch}{1.4}
\centering
\begin{longtable}{|M{0.17\linewidth}|M{0.69\linewidth  }|}
\hline
\multicolumn{2}{|c|}{} \\ 
\multicolumn{2}{|c|}{\textit{Static techniques}} \\ 
\multicolumn{2}{|c|}{} \\  \hline 
Control graph flow analysis \cite{Bodei2005Checking} & Analyze the graphical representation of all the paths that could be traversed during the execution of a program to discover security issues. \\ \hline
%Symbolic semantic graph &  Analyze the graphical representation of the program by inspecting the semantic relationships between its parts. \\ \hline 
Context-sensitive data flow analysis \cite{Sampaio2016Exploring} & Define a set of ``contexts'', i.e., scenarios in which a function can be called within the program, and analyze the data flow when the context changes. \\ \hline
Code property graph \cite{6956589} & Realize a graph-based representation of the analyzed program that can be searched for obtaining relevant information and potentially discover vulnerabilities.\\ \hline
State machine representation \cite{Lee2018Vulnerability} & A theoretical model based on program states is used to represent the behavior of the program by potentially revealing security flaws. \\ \hline

\multicolumn{2}{|c|}{} \\ 
\multicolumn{2}{|c|}{\textit{Dynamic techniques}} \\
\multicolumn{2}{|c|}{} \\  \hline
Black-box fuzzer \cite{Alsaedi2021} & Send random inputs to trigger crashes and detect vulnerabilities. No knowledge about the application's internal structure. \\ \hline

White-box fuzzing~\cite{CHEN2018118} & Perform fuzzing with full knowledge of the internal structure of the application under test. \\ \hline 

Grey-box fuzzer~\cite{CHEN2018118} & Combine black-box and white-box approaches to leverage both of them. \\ \hline
Symbolic execution \cite{Coward1988} & Analyze the program to determine through symbolic variables the inputs that trigger the execution of a specific part of a program. \\ \hline

Concolic execution~\cite{SABBAGHI2020103444} & Like symbolic execution, but using concrete values, by increasing the path exploration space. \\ \hline 

Tracing and control-data flow \cite{Foreman1993A} &  Monitor the program's control flow and data flow and keep track of triggered vulnerabilities during the execution.  \\ \hline

Bytecode instrumentation and run-time validation~\cite{10.1007/978-3-642-03007-9_9} & Modify bytecode at runtime in order to insert additional checks that can be validated to identify security flaws.  \\ \hline 

\caption{Static and dynamic techniques to discover vulnerabilities}
\label{tab:security-approaches}
\end{longtable}

\egroup

\subsection{Smart contracts}

This section provides context information useful to follow all the considerations exposed in Sections~\ref{sec:security-review} and~\ref{sec:discussion}. A common use case for WebAssembly is its application for the implementation of smart contracts. This stems from the distributed nature of smart contracts that fits perfectly with the portability and versatility of WebAssembly. Smart contracts are digital contracts stored on a blockchain, which are executed under a specific set of satisfied conditions. They are inherently related to blockchains, i.e., distributed, shared, and immutable ledgers~\cite{blockchain}.

% DEFINE Smart contract, blockchains

% Ethereum 
One of the most widely deployed blockchain platforms is Ethereum \cite{wood2014ethereum}, which extends the capabilities of the blockchain technology beyond simple financial transactions. It allows developers to create smart contracts as complex applications with self-executing code. These smart contracts get automatically executed if and when specific conditions are met, hence providing transparency, immutability, and trust in transactions. Ethereum also introduced its native cryptocurrency, known as \emph{Ether} (ETH), which is used to compensate network participants for performing computations and/or validating transactions. 
The Ethereum documentation \cite{ether-glossary} provides several relevant concepts for understanding Blockchain vulnerabilities:
\begin{itemize}
    \item \emph{Ethereum Account}: a digital identity on the Ethereum blockchain, allowing users to send and receive Ether, as well as interact with smart contracts;
    \item \emph{Address}: a unique identifier used for receiving tokens, used to identify the Ethereum Account;
    \item \emph{Block}: the data structure where the transactions or digital actions are stored;
    \item \emph{Blockchain}: a database of transactions, duplicated or shared on all computers in the network, ensuring that data cannot be altered retroactively~\cite{ether-blockchain};
    \item \emph{Smart contract}: a program that automatically executes agreements on a blockchain, like a self-enforcing digital contract~\cite{ether-smart-contracts};
    \item \emph{Transaction}: data committed to the Ethereum Blockchain signed by an originating account, targeting a specific address;
    \item \emph{Contract account}: an account containing code that executes whenever it receives a transaction from another account;
    \item \emph{Ethereum Virtual Machine (EVM)}: a stack-based virtual machine that executes bytecode;
    \item \emph{Externally Owned Account (EOA)}: the most common type of Ethereum account, controlled by a person through private keys;
    \item \emph{Internal transaction}: a transaction sent from a contract account to another contract account or an EOA;
    \item \emph{Message}: an internal transaction thet is never serialized and exists only in the Ethereum execution environment (i.e., a message is like a transaction, except it is produced by a contract and not an external actor);
    \item \emph{Message call}: the act of passing a message from one account to another;
    \item \emph{Delegatecall}: a message call that uses data of the calling contract;
    \item \emph{Receipt}: data returned by an Ethereum client to represent the result of a particular transaction, including a hash of the transaction, its block number, the amount of gas used, and, in case of deployment of a smart contract, the address of the contract.
\end{itemize}

% https://entethalliance.org/specs/ethtrust-sl/

Earlier blockchain platforms suffered from performance issues. For this reason, researchers have proposed different consensus algorithms, such as Proof-of-Share (PoS) and Delegated Proof-of-Stake (DPoS). One of the most relevant DPoS platforms is EOSIO~\cite{eos}. The EOSIO smart contract is composed of several concepts:

\begin{itemize}
    \item \emph{action}, that represents a single operation for communicating between a smart contract and an account;
    \item \emph{transaction}, that is composed of several actions;
    \item \emph{apply function}, that dispatches action handlers for validating smart contracts. The function has a `code' parameter that is used to identify the account that authorized the contract;
    \item \emph{eosio.token contract}, a standard contract that can be used to transfer tokens with the ``transfer'' functionality; 
    \item \emph{dApp}, a decentralized application that runs on a decentralized network.
\end{itemize}

\subsubsection{Vulnerabilities in smart contracts}

Smart contracts are affected by vulnerabilities that are strongly related to the smart contract domain and to the adopted blockchain platform. He et al.~(2022)~\cite{he2022survey} provide a comprehensive survey of attacks and vulnerabilities in EOSIO systems. Table~\ref{tab:relevant-vulnerabilities-smart} describes the most relevant vulnerabilities affecting smart controls.

\bgroup
\renewcommand{\arraystretch}{1.5}
\centering
\begin{longtable}{|M{0.2\linewidth}|M{0.7\linewidth}|}
\hline
\textbf{Vulnerability}                   & \textbf{Description}                                                                                                                                                                                                                                                                                          \\ \hline
Fake EOS Transfer               & An improper apply function implementation does not verify the code parameter properly, hence allowing the execution of code logic under fake EOS tokens.                                                                                                                                                   \\ \hline
Fake EOS Receipt                & If the notification of a smart contract allows unsecured relays, an attacker may invoke a transfer in \texttt{eosio.token}, notify an accomplice who relays the notification to the victim that will pass EOS checks by tricking the victim into providing services to the attacker, such as transferring tokens. \\ \hline
Fake EOS Notice                 & A variant of Fake EOS Transfer that exploits other parameters even if the system provides a valid \texttt{apply} implementation. \\ \hline
Insecure Message Call           & A message call where the authorization is not properly checked.     \\ \hline
Greedy Smart Contract           & A greedy smart contract can receive ethers, but it contains no functions to send ethers out. \\ \hline
Dangerous Delegatecall      & Attacker manipulates \texttt{msg.data} by changing the function called by the victim's contract. \\ \hline 
Block Information Dependency    & The improper utilization of a block timestamp or block number for determining a critical operation. As these variables can be manipulated by miners, they cannot be considered as reliable sources. \\ \hline
Mishandled Exceptions & Improper exception handling can cause inconsistencies during a transaction. \\ \hline
Reentrancy Vulnerability        & Invoke in a reentrant manner a function that was not designed to be reentrant\footnote{A reentrant function is a function designed to be called recursively, or by two or more processes}. \\ \hline
Rollback                        & The attacker exploits an unsafe rollback operation on a blockchain system by conditionally reverting smart contracts. Commonly used in gambling dApps.                                   \\ \hline
Missing Permission Check        & Do not properly validate permissions of sensitive operations.                                \\ \hline
Pseudo Random Number Generators & Sensitive operations that trust the insecure implementation of pseudo-random number generators can cause security issues.                                    \\ \hline
Suicidal                        & Unprotected interface allows the destruction of a victim contract.                                                                             \\ \hline
Call injection                  & An attacker is able to call a sensitive function. \\ \hline

\caption{Most relevant vulnerabilities in smart contracts.}
\label{tab:relevant-vulnerabilities-smart}
\end{longtable}
\egroup

Section~\ref{sec:attack-detection} shows the techniques that have been devised to detect this kind of attacks. 

\subsection{Trusted Execution Environments and security enclaves}

As it will be shown in Section~\ref{sec:use-cases}, several WebAssembly use-cases and enhancements entail an integration with processor security features. 
A trusted execution environment (TEE) is a hardened memory section of the processor aimed at protecting the confidentiality and integrity of both code and data~\cite{7345265}. 
One of the most utilized technologies that implement TEEs is the Intel Security Guard Extension (SGX)~\cite{intelsgx}. SGX defines the ``security enclave'' concept, i.e., a secure area where sensitive data and code can be executed. The approach offers scalability by ensuring the execution of an entire application or just a single function in the enclave. Developers can use an API to build secure applications based on Intel SGX. Intel also provides a Software Development Kit (SDK) to handle security enclaves in their code.

\section{Literature review process}
\label{sec:search}
This section illustrates the process used to review the literature on security for WebAssembly. 
The review has  followed a structured process inspired by the methodologies outlined in two relevant works: Webster and Watson (2022)~\cite{a189c9c4-218f-3e28-9a79-dae478874c48} and Kitchemann and Brereton (2013)~\cite{KITCHENHAM20132049}. 
The process proceeded through the following steps: 
\begin{enumerate}
    \item \emph{Automatic search}: We performed an automatic search with specific keywords; 
    \item \emph{Cleaning}: we removed duplicates obtained from the automatic search, and excluded out-of-scope works that did not satisfy the inclusion and exclusion criteria; 
    \item \emph{Backward and forward references review}: for each included study, we analyzed both the citations and references to identify other relevant works; 
    \item \emph{Works classification}: we performed a first examination to identify the primary contributions of each study and divided them into categories, as described in Table~\ref{tab:categories}; 
    \item \emph{Works review}: we reviewed the retrieved works in detail and developed the concepts presented in Sections~\ref{sec:security-review} and \ref{sec:discussion}. 
\end{enumerate}

\begin{table}[ht]
\centering
\begin{tabular}{|M{4cm}|M{6,5cm}|M{1,5cm}|}
\hline
\textbf{Category} & \textbf{Description} & \textbf{Nr. of studies} \\ \hline
Security Analysis   & Analyze the security of the WebAssembly ecosystem & \secanalysisworks    \\ \hline
Attack Scenarios   & WebAssembly capabilities for realizing attack scenarios & \empiricalworks     \\ \hline
Use case in cybersecurity & Leverage WebAssembly to realize a use case in the cybersecurity domain & \usecaseworks     \\ \hline
Vulnerability Discovery   & Discover vulnerabilities in WebAssembly binaries &  \vulndiscworks    \\ \hline
Security Enhancements  & Extend WebAssembly language and components to increase security & \enhancementworks \\ \hline
Others  & Relevant works to extend the literature of security works in WebAssembly & \otherworks     \\ \hline

\end{tabular}
\caption{Security categories for the analyzed literature}
\label{tab:categories}
\end{table}

Table~\ref{tab:process-works} reports the overall number of publications that we reviewed at each step of the aforementioned process. 

\begin{table}[ht!]
\renewcommand{\arraystretch}{1.2}
\begin{tabular}{|M{5cm}|M{3,5cm}M{3,5cm}|} \hline
 \textbf{Keywords} & \textbf{Nr. of papers from SCOPUS}  & \textbf{Nr. of papers from IeeeXplore} \\ \hline
``wasm security''             & 43                     & 20                         \\ \hline
``wasm vulnerability''        & 18                     & 8                          \\ \hline
``webassembly security''      & 91                    & 44                         \\ \hline
``webassembly vulnerability'' & 26                     & 10                         \\ \hline
                     & 178                    & 82      \\ \hline                
\textit{Merged and without duplicates} &  \multicolumn{2}{| c |}{114} \\ \hline
\textit{After inclusion and exclusion criteria} &  \multicolumn{2}{| c |}{62} \\ \hline
\textit{After reference analysis} &  \multicolumn{2}{| c |}{\textbf{\totalworks}} \\ \hline

\end{tabular}
\caption{Publications analyzed at each step of the literature review process}
\label{tab:process-works}
\end{table}

\subsection{Automatic search}
We search for four relevant keyword pairs that relate security to WebAssembly: (i) ``webassembly security''; (ii) ``webassembly vulnerability''; (iii) ``wasm security''; (iv) ``wasm vulnerability''. 

We exclude other keywords, since a quick analysis shows that they basically provide almost equal results to the utilized keyword combinations. For instance, ``cybersecurity'' would provide the same results as ``security''. The same holds  true for the use of plural forms such as ``vulnerabilities'' instead of ``vulnerability''. 

We use two relevant platforms containing peer-reviewed articles, i.e., IEEE Xplore~\cite{ieeexplore} and Scopus~\cite{scopus}.
After obtaining the works, we merge them and remove duplicates. Then, we apply the inclusion and exclusion criteria described further below, to select just in-scope works. Finally, we perform the backward and forward reference investigation to find other relevant works.

\subsection{Works' inclusion and exclusion criteria}

We only include works specifically focused on the security domain. However, we include into a secondary category named ``others'' those works that we consider essential sources for investigating new security research topics in WebAssembly, even if they are not specifically focused on security.
We primarily consider peer-reviewed articles, but during the backward and forward reference review, we also include relevant literature from pre-print services~\cite{arXiV}, as well as relevant technical papers from highly recognized international cyber-security conferences~\cite{blackhat}.
%Furthermore, we have decided to include works that are not explicitly focused on cybersecurity but that bring relevant contributions to starting future works on cybersecurity.
We exclude works not written in English and mention in Section~\ref{sec:other} works that explore security topics but do not provide satisfactory analyses, such as preliminary studies and poster papers. 

\subsection{Works' classification}

After the document retrieval phase, we first analyze each work's objective to categorize and determine its primary contribution. We delineate several categories and organize items based on their contribution type. Then, we go deeper into each paper to scrutinize its technical details or gain a better understanding of any proposed case studies by exploring the features of other works. Finally, we synthesize our analysis, as detailed in Section~\ref{sec:security-review}.
For the specific category of malware detection, we exclude studies that discover vulnerabilities without involving WebAssembly features, e.g.,  crypto mining detection through machine-learning models on network traffic~\cite{10389593, 10.1145/3570361.3613283}, or dynamic analysis focused on binaries not specifically generated by WebAssembly code~\cite{darabian2020}. In order to classify the reviewed works, we identify six main categories:

\begin{itemize}

\item \emph{Security Analysis}: papers that analyze the security of the WebAssembly ecosystem. These works are useful in providing a preliminary context of security flaws that affect WebAssembly programs;
\item \emph{Empirical studies}: papers that investigate security aspects of WebAssembly by analyzing real-world scenarios. These works apport relevant empirical contributions;
\item \emph{Attack Scenarios}: these papers illustrate approaches that leverage WebAssembly capabilities for conducting complex attacks;
\item \emph{Vulnerability Discovery}: the contribution of these papers is to provide approaches to discover vulnerabilities in applications based on WebAssembly;
\item \emph{Security Enhancements}: in these papers, authors aim to extend the WebAssembly components in order to reduce security flaws that are examined in security analysis works; 
\item \emph{Use cases in security}: in these papers, authors leverage WebAssembly features to increase the security aspects of a proposed solution, system, and/or framework;
\item \emph{Others}: some relevant papers are not specifically focused on cybersecurity but can be considered as preliminary studies representing ``cornerstones'' for conducting novel security-related research activities. 
\end{itemize}

\section{Related Works}
\label{sec:related}
Several works offer pertinent reviews on WebAssembly that do not explicitly center on security. 

WebAssembly runtimes are analyzed by Wang and Wenwen (2022)~\cite{9975423}. In their study, the authors review five predominant standalone WebAssembly runtimes and construct a benchmark suite for comparative purposes. They assert that performance is influenced by various factors and outline criteria for selecting the best runtime based on specific use cases.  Additionally, they offer insights into enhancing runtime performance by including crucial information into compiled WebAssembly binaries. 

%Although it is not a comprehensive review, the work proposed by Wang and Wenwen~\cite{9975423} provides a characterization study of five predominant WebAssembly runtimes and obtains noteworthy results that can be examined for improving them.

In the realm of cloud-edge computing~\cite{9083958}, Kakati et al. (2023)~\cite{10205816} provide a comprehensive review of the adoption of WebAssembly.

The Internet of Things (IoT) is another relevant context where WebAssembly is widely adopted~\cite{Schoder2018}. Ray and Partha Pratim (2023)~\cite{Ray2023} provide a comprehensive survey about the adoption of WebAssembly in IoT systems by illustrating the potential of WebAssembly characteristics, providing a valuable resource for developers through a comparison of WebAssembly tools and toolchains, and illustrating future insights for continuing this research field. 

\subsection{Security studies}

Section~\ref{sec:sec-analysis} delves into related works offering security analysis for WebAssembly compilers, runtimes, and the impacts of porting programs written in unsafe memory languages~\cite{Lehmann2020217, Stievenart2021132, Stiévenart20221713, bhat-chasm}. Notably, Kim et al. (2022)~\cite{9860829} and Harnes and Morrison (2024)~\cite{Harnes2024} provide a comprehensive survey on techniques and methods for WebAssembly binary security. They also propose future research directions to address open problems identified in their studies, such as protection from malicious WebAssembly binaries and hardening the inherent security of WebAssembly. 

%Our study encompasses the works reviewed by the above mentioned authors as well as additional literature. Specifically, we broaden the scope of covered security domains by examining other types of works, as detailed in Section~\ref{sec:security-review}.

Tables~\ref{tab:compare-related}, \ref{tab:compare-related-two}, and \ref{tab:total-number} present a comparative analysis of reviews. Our study extends beyond the scope of previous works by encompassing various types of literature, including those focused on enhancing the security of WebAssembly components, utilizing WebAssembly for security purposes in various use cases, conducting empirical studies, and performing security analyses. Furthermore, we cover five additional categories, culminating in the analysis of \totalworks~works. 

\begin{table}[ht!]
\centering
\begin{tabular}{M{0.4\linewidth}|M{0.15\linewidth}|M{0.15\linewidth}|M{0.1\linewidth}|}
                          & Kim et al. (2022) \cite{9860829} & Harnes and Morrison (2024) \cite{Harnes2024} &\textbf{Our work} \\ \hline
Security analysis         &  &   &    \cmark      \\ \hline
Security enhancement      & \cmark    &   & \cmark          \\ \hline
Vulnerability discovery   &  \cmark   &    \cmark &     \cmark     \\ \hline
Use case in cybersecurity &      &    &    \cmark      \\ \hline
Attack scenarios &      &    &    \cmark      \\ \hline
Attacks' detection        & \cmark     &   \cmark & \cmark          \\ \hline
Empirical studies & \cmark    &    & \cmark          \\ \hline
Other works               &   &    &       \cmark   \\ \hline
\end{tabular}
\caption{Comparison of studies based on our classification and related ones. Our review encompasses additional research by incorporating other types that address both WebAssembly and security.}
\label{tab:compare-related}
\end{table}

\begin{table}[ht!]
\centering
\renewcommand{\arraystretch}{1.2}
\begin{tabular}{M{0.3\linewidth}|M{0.2\linewidth}|M{0.2\linewidth}|M{0.1\linewidth}|}
                          & Kim et al. (2022) \cite{9860829} & Harnes and Morrison (2024) \cite{Harnes2024} &\textbf{Our work} \\ \hline
Vulnerability discovery   & 7   & 15    &  \textbf{\vulndiscworks}   \\ \hline
Attacks' detection & 9 & 7   &  \textbf{\attackdetectionworks}   \\ \hline
Security enhancement & 8 & 0 & \textbf{\enhancementworks} \\ \hline
\end{tabular}
\caption{Number of analyzed publications for each category that is covered by related works}
\label{tab:total-number}
\end{table}

\begin{table}[ht!]
\centering
\renewcommand{\arraystretch}{1.2}
\begin{tabular}{M{0.3\linewidth}|M{0.4\linewidth}|M{0.2\linewidth}}
Kim et al. (2022) \cite{9860829} & Harnes and Morrison (2024) \cite{Harnes2024} &\textbf{Our work} \\ \hline
24 & 22 & \textbf{\totalworks}
\end{tabular}
\caption{Total number of reviewed works. }
\label{tab:compare-related-two}
\end{table}

It is worth noting that we classified certain publications (\cite{wassail2021, CabreraArteaga2021}) under the ``other'' category instead of grouping them with related studies. Our decision was based on our analysis, which found that these publications lacked sufficient information to be classified as vulnerability discovery efforts.

To the best of our knowledge, this is the first work providing a comprehensive review of security research in the WebAssembly domain.

\newpage
\section{WebAssembly and Security: a review}
\label{sec:security-review}

This section presents the works analyzed in this study.
As previously outlined, Table~\ref{tab:categories} summarizes the proposed categories detailed in Section~\ref{sec:search}, and displays the number of analyzed studies for each category.

\subsection{Security Analysis}
\label{sec:sec-analysis}

Although WebAssembly has been designed with security in mind, relevant works underline that it can suffer from security issues. One significant study conducted by Lehmann et al. (2020)~\cite{Lehmann2020217} highlights this concern. In their research, the authors demonstrate that certain classic vulnerabilities, which are mitigated by compiler system enhancements in native binaries, remain exploitable in WebAssembly. They conduct an extensive security analysis of  WebAssembly's linear memory system, present examples of vulnerable applications, outline attack primitives, demonstrate end-to-end exploits for identified vulnerabilities, and propose possible mitigation strategies to harden WebAssembly binaries. 
In particular, the authors analyze the impact of classic attack primitives and categorize them into three main types: write operations (including stack-based buffer overflow~\cite{stack}, stack overflows, and heap metadata corruptions), data overwrites (such as overwriting stack and heap variables, as well as ``constant'' data), and triggers for unexpected behavior (such as redirecting indirect calls, injecting code into the host environment, and overwriting application-specific data). They illustrate how these attack primitives can be combined to execute end-to-end attacks such as cross-site scripting, remote code execution, and arbitrary file writing. 

To mitigate the analyzed security flaws, the authors propose three possible approaches: extending the WebAssembly language, enhancing compilers and toolchains, and developing robust and safe libraries for use during development. 

Stievenart et al. (2021)~\cite{Stievenart2021132} explore the security vulnerabilities stemming from the absence of protections in WebAssembly compilers. In their study, the authors compile $4.469$ C programs containing known buffer overflow vulnerabilities (both stack-based and heap-based) to x86 code and WebAssembly. They reveal that $1.088$ programs exhibit a different behavior when compiled to WebAssembly. In particular, WebAssembly is found to be more vulnerable to stack-smashing attacks due to the absence of security measures like stack canaries. 

As illustrated before, memory management in low-level programs is the primary cause of security flaws. Therefore, the most affected programming language is C~\cite{10.5555/576122}, as it provides the lowest abstraction level with respect to the underlying machine so that it can fully interact with the hardware.

%&but does not natively support memory safety controls such as Rust. 
Stiévenart et al. (2022) \cite{Stiévenart20221713} investigate the security implications of cross-compiling C programs to WebAssembly. 
In their study, the authors compile and execute $17.802$ programs with common vulnerabilities for both 64-bit x86 and WebAssembly. They find that $4.911$ binaries yield different results due to three main reasons: the absence of security measures in WebAssembly, different semantics of execution environments, and variations in used standard libraries. Critical differences affecting security include the implementation of \texttt{malloc()} and \texttt{free()} functions, and the lack of both stack-smashing and memory protection mechanisms. 
To address these issues, the authors suggest extending WebAssembly to incorporate memory semantics in C and C++ code. They also emphasize the importance of static analysis in identifying security flaws and highlight related works aiming to enhance the WebAssembly runtime and mitigate the security impacts~\cite{Menetrey2021205, Nieke202113}.

One of the noteworthy resources for developers is the technical study conducted by Brian McFadden et al. (2018)~\cite{bhat-chasm}. In this work, the authors delve into various facets of WebAssembly security. In particular, they: 
\begin{itemize}
    \item  analyze Emscripten's implementation of compiler and linker-level exploit mitigation; 
    \item  introduce new attack vectors and methods of exploitation in WebAssembly, demonstrating that buffer overflow or indirect function calls can lead to cross-site scripting or remote code execution attacks; 
    \item provide examples of memory corruption exploits in the WebAssembly environment;
    \item outline best practices and security considerations for developers aiming to integrate WebAssembly into their products. 
\end{itemize}

\subsection{Empirical Studies}
These works present empirical case studies to analyze a specific phenomenon in the WebAssembly ecosystem.
Early studies analyze the diffusion of WebAssembly for executing cryptojacking attacks in the wild.

R\"{u}th et al. (2018)~\cite{10.1145/3278532.3278539} introduce a new fingerprinting method aimed at identifying miners, which revealed a factor of $5,7$ more miners compared to publicly available
block lists. They also classify mining websites among $138$ million domains sourced from the largest top-level domains and the Alexa top $1$ million websites~\cite{alexa}. The authors demonstrate that the prevalence of browser mining is low, at less than $0,08\%$ and that $75\%$ of the mining sites utilize Coinhive, a web-based mining provider. %In their study, R\"{u}th et al. (2018) \cite{10.1145/3278532.3278539} come to similar conclusions by observing that in the Alexa Top 1 Million dataset 1 out of 500 sites hosting a mining script, and the landscape of cryptojacking miners is dominated by variants of Coinhive. 

Musch et al. (2019)~\cite{10.1007/978-3-030-22038-9_2, Musch2019} analyze websites from Alexa top one million ranking~\cite{alexa} in detail and find that $1$ out of $600$ websites uses WebAssembly. However, they reveal that 50\% of all sites using WebAssembly employ it for mining and obfuscation purposes. 

%The cryptojacking is also explored by Konoth et al. (2018) \cite{10.1145/3243734.3243858}, which analyzes attacks and countermeasures on 28 Coinhive-like services widely used by mining websites.
Hilbig et al. (2021)~\cite{10.1145/3442381.3450138} also analyze the usage of WebAssembly worldwide, presenting different results.
They collect $8.461$ WebAssembly binaries from a wide range of sources and conclude that:
\begin{itemize}
    \item $64,2\%$ of the binaries are compiled from memory-unsafe languages, such as C and C++;
    \item less than $1\%$ of all binaries are related to crypto mining, with a greater diffusion of game-based, text processing, visualization, and media applications;
    \item $29\%$ of all binaries are minified, suggesting a need for approaches to decompile and reverse engineer WebAssembly.
\end{itemize}

Romano et al. (2021)~\cite{9678776} conduct two empirical studies to perform a qualitative analysis of $146$ bugs related to WebAssembly in Emscripten and a quantitative analysis of $1.054$ bugs in three open-source compilers, namely, AssemblyScripts, Emscripten, and Rustc/Wasm-Bindgen. They draw the conclusion that sensitive information can be disclosed in Trusted Execution  Environments (TEE). Puddu et al. (2022)~\cite{puddu2022the} analyze the impacts of executing confidential code on top of a  WebAssembly runtime within a TEE and show that using an intermediate representation such as  WebAssembly can lead to code leakage.

\subsection{Use cases in cybersecurity}
\label{sec:use-cases}
WebAssembly is extensively used in heterogeneous domains where security is a crucial non-functional requirement. 
%Table~\ref{tab:use-cases} summarizes the authors' contributions by underlying the criteria for selecting WebAssembly as a core environment for their research avenues.
The following subsections summarize the most interesting contributions we found in the literature, by underlying the main research paths they have tried to follow.

\subsubsection{Browser extensions}
% BROWSER
WebAssembly was designed to provide an assembly language for the web,  especially for modern browsers. Therefore, there is extensive literature about realizing browser extensions with this technology. 
As WebAssembly is a portable language, all the works presented in this section could potentially be ported toward browser environments. However, the works that we classify as ``Browser extensions''  have specifically been realized to extend browser capabilities and increase portability, isolation, and efficiency. 

Baumgärtner et al. (2019)~\cite{Baumgärtner2019} present a Disruption-tolerant Networking (DTN) system for browsers that leverages WebAssembly and the Bundle Protocol $7$ draft implementation~\cite{rfc9171}. 
Narayan et al. (2020)~\cite{Narayan2020699} define a new framework (RLBox) that implements fine-grained isolation for sandboxing third-party libraries. The framework has been integrated into production Firefox to sandbox the libGraphite font shaping library. 
Chen et al. (2023)~\cite{10.4018/IJSWIS.334591} propose a novel approach for digital rights management of streaming. The work allows users to easily fetch videos with a plug-and-play approach and enhances decryption efficiency by minimizing interactions with the server. 

WebAssembly in the browser is employed by Jang et al. (2022)~\cite{10.1145/3545948.3545971} to realize a distributed public collaborative fuzzing system for the security of applications emulated inside the browser engine. 

% CRYPTO, TEE and PRIVACY
\subsubsection{Cryptography, Trusted Execution Environments and Privacy}

Various works aim to safeguard data integrity and confidentiality through cryptographic techniques, Trusted Execution Environments, and privacy-preserving methods. These approaches are tailored to specific domains and can be implemented in hardware, embedded systems, or browsers. 
Sun et al. (2020)~\cite{Sun202201871} leverage WebAssembly and the Web Cryptography API to devise a client-side encrypted storage system capable of storing and sharing data across heterogeneous platforms. 
Attrapadung et al. (2018)~\cite{Attrapadung2018} use WebAssembly to implement an efficient two-level homomorphic public-key encryption within prime-order bilinear groups. 
Seo et al. (2023)~\cite{Seo20232091} introduce a portable and efficient WebAssembly implementation of the Crystals-Kyber post-quantum Key Encapsulation Mechanism~\cite{crystals}.  
Zhao et al. (2023)~\cite{Zhao20234015} define a novel approach for constructing reusable, rapid, and secure enclaves. They introduce three techniques, namely, enclave snapshot and rewinding, nested attestation, and multi-layer intra-enclave compartmentalization, for enabling rapid enclave reset and robust security.
WebAssembly is used to realize publish and subscribe environments for cloud-edge computing. Both Almstedt et al. (2023)~\cite{Almstedt2023} and  M\'{e}n\'{e}trey et al. (2024)~\cite{menetrey2023} design a novel publish and subscribe middleware running in Intel SGX for trustworthy and distributed communications. Qiang et al. (2018)~\cite{Qiang2018451} utilize the sandbox properties of WebAssembly combined with Intel SGX enclaves to develop a two-way sandbox. This approach offers dual protection: safeguarding sensitive data from potential threats posed by the cloud provider and shielding the host runtime from potential attacks initiated by malicious users.
Other works show that WebAssembly can be used to implement a performant, memory-safe, and portable client-side-hashing library (Riera et al. (2023)~\cite{icissp23}), or even a fully-fledged client-side storage platform (Sun et al. (2022)~\cite{Sun202201871}).

% IAM 
\subsubsection{Identity and Access Management}

Another integration of Intel SGX and WebAssembly is provided by Goltzsche et al. (2019)~\cite{10.1145/3361525.3361541}. In their work, the authors present AccTEE, a framework for realizing fine-grained permissions for resource accounting by preserving the confidentiality and integrity of code and data.
Also, Zhang et al. (2021)~\cite{Zhang2021} propose EL PASSO, a portable and ``zero-configuration'' module to allow the setup and configuration of privacy-preserving single-sign-on operations. 

% CLOUD AND CLOUD-EDGE COMPUTING
\subsubsection{Cloud, cloud-edge computing and IoT}

Cloud and cloud-edge computing are common use cases of WebAssembly adoption \cite{10365297}. Edgedancer~\cite{Nieke202113} is a platform for supporting portable, provider-independent, and secure migration of edge services. Ménétrey et al. (2022)~\cite{Ménétrey20223} envisage that WebAssembly with TEEs is a promising combination to realize cloud-edge continuum~\cite{10.1007/978-3-031-36889-9_16}.

Sun et al. (2022)~\cite{Sun202201871} primarily focus on the browser domain as a platform aimed at realizing a secure cloud storage environment. This subject is also addressed by Song et al. (2023)~\cite{10098652}, who provide a cross-platform secure storage architecture through the realization of a new primitive called Controllable Outsourced Attribute-Based Proxy Re-Encryption (COAB-PRE) alongside the cross-platform capabilities of WebAssembly. 
Although WebAssembly provides several security benefits,  performance overheads should be investigated when it is adopted in contexts with constrained resources, such as IoT and edge devices. Wen et al. (2020)~\cite{Wen2020353} address this problem by defining a novel operating system that integrates Rust and WebAssembly to make IoT applications more efficient and secure. A similar approach is also proposed by Li and Sato (2023)~\cite{Li202342}, who implement a kernel focused on isolation, security, and customizability through the combination of WebAssembly and Rust.

\subsubsection{Face recognition systems}

Preliminary work has been conducted on using WebAssembly to realize face recognition systems. Indeed, WebAssembly ensures a client-side solution that preserves image privacy and could potentially reduce the costs of adopting external services. 
Pillay et al. (2019)~\cite{Pillay2019} provide a real-time browser-based application that is able to attain a facial recognition rate of $91.67\%$, while Mart{\'\i}n Manso et al. (2021)~\cite{10.1007/978-3-030-92325-9_7} compare JavaScript (CPU), WebGL and WebAssembly for face detection, and conclude that WebAssembly strikes a good balance in terms of detection accuracy and time overheads.

\subsection{Attack scenarios}
Another research field deals with innovative approaches for conducting attacks against WebAssembly programs. Most articles fall under two primary categories:
\begin{itemize}
    \item \emph{Side-channel attacks}: these works demonstrate how it is possible to conduct side-channel attacks to disclose sensitive keys or realize covert channels~\cite{cryptoeprint:2018/119, 277134, 10.1145/3488932.3517411, Katzman20231955};
    \item \emph{Evasion techniques for malware detection}: these works illustrate in which way it is possible to bypass malware detection systems through WebAssembly  \cite{CABRERAARTEAGA2023103296, 9833626, 10.1007/978-3-031-44245-2_8, 10.1007/978-3-031-35504-2_4}. 
\end{itemize}

\begin{table}[ht!]
\renewcommand{\arraystretch}{1.3}
\centering
\begin{tabular}{|c|c|}  \hline
\textbf{Cite}                                                 & \textbf{Type of implemented attack} \\ \hline
Genkin et al. (2018) \cite{cryptoeprint:2018/119}             & Side-channel attacks                 \\ \hline
Easdon et al. (2022) \cite{277134}                            & Side-channel attacks                \\ \hline
Rokicki et al. (2022) \cite{10.1145/3488932.3517411}         & Side-channel attacks                \\ \hline
Katzman et al. (2023) \cite{Katzman20231955}                  & Side-channel attacks (cache)     \\ \hline 
Cabrera-Arteaga et al. (2023) \cite{CABRERAARTEAGA2023103296} & Malware detection evasion           \\ \hline
Romano et al. (2022) \cite{9833626}  & Malware detection evasion \\ \hline 
Cao et al. (2023) \cite{10.1007/978-3-031-44245-2_8} & Malware detection evasion \\ \hline
Loose et al. (2023) \cite{10.1007/978-3-031-35504-2_4}        & Malware detection evasion           \\ \hline
Oz et al. (2023) \cite{291205} & Ransomware attack \\ \hline

\end{tabular}
\caption{Works focused on describing attack examples in WebAssembly.}
\label{tab:attack-scenarios}
\end{table}

Oz et al. (2023) \cite{291205} provide a divergent work by showing how it is possible to leverage WebAssembly to implement ransomware attacks \cite{10.1007/978-981-13-1274-8_31}.
Table~\ref{tab:attack-scenarios} summarizes the above information.

\subsubsection{Side-channel attacks}

Genkin et al. (2018) \cite{cryptoeprint:2018/119} utilize WebAssembly to induce CPU cache contention and extract sensitive data from other programs' memory accesses. They demonstrate that this technique can extract sensitive keys from famous cryptographic libraries.
An innovative cache-based attack is also illustrated in the work of Katzman et al. (2023)~\cite{Katzman20231955}, where authors conclude that transient execution can increase the effectiveness of cache attacks.

Easdon et al. (2022)~\cite{277134} leverage WebAssembly features to demonstrate a novel technique for prototyping microarchitectural attacks. They achieve this by implementing two open-source frameworks, libtea and SCFirefox, and developing proof-of-concepts for executing Foreshadow~\cite{vanbulck2018foreshadow} and Load Value Injection (LVI)~\cite{vanbulck2020lvi} attacks.

Rokicki et al. (2022)~\cite{10.1145/3488932.3517411} introduce the first port contention side-channel attack conducted entirely within the browser environment. Their study demonstrates that certain WebAssembly instructions can induce port contention attacks, enabling the extraction of sensitive data from concurrent processes executing on the same CPU core. 

\subsubsection{Malware detection evasion}

The other two works propose innovative techniques to evade malware detection in WebAssembly binaries. 
Cabrera-Arteaga et al. (2023)~\cite{CABRERAARTEAGA2023103296} analyze the potential of WebAssembly binary diversification to circumvent detection by antivirus systems. On the other hand, Loose et al. (2023)~\cite{10.1007/978-3-031-35504-2_4} introduce a novel approach to embed adversarial payloads into the instruction stream of binaries, thereby bypassing malware detection systems. 

Additionally, Romano et al.~(2022) \cite{9833626} demonstrate the feasibility of obfuscating JavaScript to evade malware detection systems by converting specific JavaScript instructions into WebAssembly.
Cao et al. (2023)~\cite{10.1007/978-3-031-44245-2_8} develop a framework for static binary rewriting, which can be used for various purposes such as obfuscation. Although the primary focus is not solely on evading malware detection systems, the framework can be adapted for that purpose, as well as for patching vulnerabilities or binary instrumentation.

\subsection{Attack detection solutions}
\label{sec:attack-detection}

These works focus on detecting attacks in WebAssembly programs, primarily targeting cryptojacking attacks. As shown in Table~\ref{tab:detected-attacks}, current research is essentially focused on detecting crypto-mining activities facilitated by WebAssembly technology. 
%Table~\ref{tab:attacks-detection-summary} provides a summary of the security outcomes, comparing the benchmarks used and the results obtained across these works.

\bgroup
\renewcommand{\arraystretch}{1.5}
\begin{longtable}{|M{2cm}|M{7,5cm}|M{2,5cm}|}
\hline
\centering
\textbf{Work} & \textbf{Detection approach} & \textbf{Type of detected attacks} \\ \hline 
Konoth et al. (2018)~\cite{10.1145/3243734.3243858} & Runtime detection on the basis of three approaches discovered through static analysis of WebAssembly malicious binaries & Cryptojacking \\  \hline
Wang et al. (2018)~\cite{Wang2018122}               & At runtime, monitor execution through in-line scripts                           & Cryptojacking \\  \hline
Rodriguez and Possega (2018)~\cite{Rodriguez2018} & Analyze WebAssembly and JavaScript APIs and train a Support Vector Machine (SVM). Show the feasibility of deploying the solution as a browser extension & Cryptojacking \\ \hline
Kharraz et al. (2019)~\cite{10.1145/3308558.3313665}  & Support Vector Machine classifier & Cryptojacking \\ \hline
Bian et al. (2020)~\cite{Bian20203112}              & At runtime                           & Cryptojacking \\  \hline
Kelton et al. (2020)~\cite{Kelton2020}              & At runtime, through a Support Vector Machine (SVM) classifier                 & Cryptojacking \\  \hline
Mazaheri et al. (2020)~\cite{9261920}   & Static analysis by analyzing features of web pages that indicate the presence of time-based side-channel attacks. & Side-channel \\ \hline %(memory deduplication and Spectre attacks) 
Romano et al. (2020)~\cite{9286112}                 & Static analysis, by reconstructing binaries in ``standard'' WebAssembly format, and analyzing the generated intra-procedural and inter-procedural Control-Flow Graphs (CFGs)     & Cryptojacking \\  \hline
Yu et al. (2020)~\cite{10.1007/978-981-33-4922-3_5}  & Static analysis with machine learning and runtime detection systems. Provide both detection and blocking capabilities.  & Cryptojacking \\ \hline
Naseem et al. (2021)~\cite{Naseem2021}              & At runtime, leveraging a Convolutional Neural Network (CNN) model trained with a comprehensive dataset of current malicious and benign WebAssembly binaries                           & Cryptojacking \\  \hline
Tommasi et al. (2022)~\cite{Tommasi2022} & At runtime, browser extension based on a Support Vector Machine (SVM) classifier & Cryptojacking \\ \hline
Breitfelder et al. (2023)~\cite{Breitfelder2023753} & Static WebAssenbly analysis that is able to capture call-, control-, and data-flow graphs for further analysis. The detection capability is one of the discussed use-cases. & Cryptojacking \\ \hline 
\caption{Works about detecting attacks against WebAssembly applications}
\label{tab:detected-attacks}
\end{longtable}
\egroup    

%%An Yavarzadeh et al. (2023) \cite{halfhalf} \\ \hline

\subsection{Security enhancements}
Several studies have extended WebAssembly components or integrated new systems to enhance their security level. 
%Table~\ref{tab:security-enhance-extension-point} illustrates which components are extended or integrated with WebAssembly to improve security aspects. 
Some authors have extended the semantics of the WebAssembly language to include protection against various threats. For instance, Watt et al. (2019)~\cite{Watt2019} propose CT-Wasm, a type-driven secure language for writing cryptographic libraries to protect against time-based side-channel attacks. Similarly, Michael et al. (2023)~\cite{Michael2023425} focus on protecting WebAssembly program memory by introducing MSWasm, a WebAssembly semantic addressing memory unsafe vulnerabilities. They introduce a new memory region called \textit{segment memory} and new instructions for interacting with it. Additionally, Geller et al. (2024)~\cite{10.1145/3632922} extend WebAssembly semantics to define indexed types~\cite{ZENGER1997147} to ensure safety properties over program values. Another language extension aimed at ensuring memory safety is proposed by Zhang et al. (2023)~\cite{Zhang2023662}, who allow the definition of canaries in the program. 

Several works modify the WebAssembly compiler to formally verify security properties~\cite{Protzenko20191256, 279990} or reduce the attack surface in embedded systems~\cite{Narayan2023266, halfhalf}. Others focus on enhancing the WebAssembly runtime environment to create a trusted execution environment that increases isolation and prevents memory attacks during execution. 

In order to protect the execution of WebAssembly programs, several authors have designed additional mechanisms. For example, Sun et al. (2019)~\cite{9047432} propose SELWasm, a code protection technique that implements a self-checking mechanism to prevent code execution on unauthorized websites, along with an ``Encryption \& Decryption'' approach for ensuring code confidentiality. Similarly, Brandl et al. (2023)~\cite{Brandl2023} present a novel interpreter for WebAssembly capable of performing complex analyses such as inter-procedural control and data-flow information. This study can be considered as a foundation for future static security analysis approaches. 

Additionally, Pop et al. (2022)~\cite{Pop202243} propose an interpreter that extends WASMI 14~\cite{wasmi} by adding pausing, serialization, deserialization, and resumption primitives for implementing a portable enclave service.

Other works explore novel approaches to obfuscate WebAssembly binaries, aiming to increase protection against reverse engineering or to highlight the adversarial capabilities of WebAssembly in evading malware detection systems. 

% Verification => Vulnerability discovery
Other works introduce verification of programs after compilation. Watt (2018)~\cite{Watt2018} mechanizes and verifies the WebAssembly specification using the theorem prover Isabelle~\cite{isabelle}, illustrating two demonstrations that implement a verified type-checker and interpreter for WebAssembly.
This work is extended by Watt et al. (2023)~\cite{10.1145/3591224}, who define a WebAssembly interpreter written in Isabelle/HOL~\cite{nipkow2002isabelle} to validate the Wasmtime interpreter~\cite{github-wasmtime}.
Johnson et al. (2021)~\cite{johnson2021} introduce VeriWasm, a static offline verifier for native x86-64 binaries compiled to WebAssembly. The verifier checks the boundaries of memory accesses, control flows, and stack usage, ensuring the satisfaction of Software Fault Isolation (SFI)~\cite{10.1145/173668.168635} in binaries.
Another area of research focuses on improving malware detection in WebAssembly binaries. Xia et al. (2024)~\cite{10.1007/978-3-031-51476-0_13} highlight the issue of emerging JavaScript-WebAssembly multilingual malware (JWMM), which reduces the effectiveness of antivirus solutions. The authors introduce a novel approach to semantically reconstruct WebAssembly binaries, generating a high-level structure aimed at enhancing antivirus detection capabilities.

Watt et al. (2021)~\cite{watt2021} extend the WebAssembly semantic and provide two different theorem provers: WasmCert-Isabelle and WasmCert-Coq.

\subsubsection{Works that enhance WebAssembly indirectly}

Some works do not extend WebAssembly components directly but address security issues by enhancing other systems. Schrammel et al. (2020)~\cite{255298} improve WebAssembly security indirectly by enhancing the JavaScript (JS) V8 engine. The authors underline that the JS engine compiles some JavaScript parts in WebAssembly and identify several potential attacks against the compiled code. They propose a JS engine extension that provides isolation mechanisms using modern CPUs to enforce domain-based access controls on memory pages. Vassena et al. (2021)~\cite{10.1145/3434330} introduce Blade, an approach to automatically eliminate speculative leaks from cryptographic code by extending ``Cranelift''~\cite{cranelift}, a WebAssembly to x86 compiler. Song et al. (2023)~\cite{10296897} propose an interesting method for blocking heap memory corruption attacks by shadowing metadata from linear memory in trusted JavaScript memory. They provide a JavaScript API that WebAssembly programs can call to shadow and validate the metadata of the used linear memory. 

\subsection{Vulnerability Discovery}
\label{sec:discovered-vuln}
Several works aim to discover vulnerabilities in WebAssembly binaries. Tables~\ref{tab:vuln-discovery-static} and \ref{tab:vuln-discovery-dynamic} categorize these efforts into two groups: (i) works that leverage static security analysis and (ii) works proposing dynamic analysis. Each approach is experimented with various techniques, allowing the discovery of different types of vulnerabilities.

Many studies focus on identifying issues in smart contracts. This clearly indicates a significant adoption of WebAssembly in the blockchain domain. 

\bgroup
\centering
\renewcommand{\arraystretch}{1.4}
\begin{longtable}{M{0.2\linewidth}|M{0.25\linewidth}|M{0.45\linewidth}}  
 \textbf{Work}                                           & \textbf{Static analysis approach}                                              & \textbf{Discovered vulnerabilities}                                                               \\ \hline
  Quan et al. (2019) \cite{quan2019evulhunter}      & Control Graph Flow analysis                                          & Fake EOS transfer, fake EOS notice                                                             \\   \hline
   Yang et al. (2020) \cite{Yang202021}                    & Symbolic Semantic Graph                                              & Integer overflow, pseudo-random number generator, insecure message call                      \\ \hline
   Li et al. (2022) \cite{Li2022746} & Context-sensitive data flow analysis & Fake EOS Transfer, Forged Transfer Notification, and Block Information Dependency \\ \hline
Brito et al. (2022) \cite{Brito2022} &  Code property graph & Buffer overflow, use after free, double free, dangerous function usage, and format strings vulnerabilities \\ \hline
                                   Tu et al. (2023) \cite{Tu2023103}                       & State Machine Representation                                         & Rollback, missing permission check, integer overflow                                           \\ \hline
\caption{Vulnerability discovery works that propose a static analysis approach}
\label{tab:vuln-discovery-static}
\end{longtable}
\egroup

\bgroup
\centering
\renewcommand{\arraystretch}{1.4}
\begin{longtable}{M{0.2\linewidth}|M{0.25\linewidth}|M{0.45\linewidth}}  
 \textbf{Work}                                           & \textbf{Dynamic analysis approach}                                              & \textbf{Detected vulnerabilities}                                                               \\ \hline
Huang et al. (2021) \cite{10.1145/3457913.3457920}   & Black-box fuzzer & Block information dependency, forged transfer notification, fake EOS transfer \\ \hline
Jiang et al. (2021) \cite{Jiang2021926}                 & Symbolic execution                                                   & Greedy, dangerous DelegateCall, block information dependency, reentrancy, mishandled exception \\ \hline
  He et al. (2021) \cite{He20211271}                      & Symbolic execution                                                   & Fake EOS, fake Receipt, rollback, missing permission check                                     \\\hline
 %Li et al. (2022) \cite{Li2022}                          & Grey-box fuzzing                                                     & Fake EOS Transfer, forged transfer notification, block information dependency                  \\\hline
                                   Daniel Lehmann et al. (2021) \cite{lehmann2021fuzzm}    & Fuzzing                                                              & Crash in binary programs, mainly related to buffer overflow conditions                        \\\hline
Marques et al. (2022) \cite{marques_et_al:LIPIcs.ECOOP.2022.11} & Concolic execution & Not specified \\ \hline
Yang et al. (2022) \cite{Yang2022322} & Tracing and control-data flow & Access control vulnerabilities  \\ \hline
Khan et al. (2023) \cite{10190488}  & Fuzzing & data races, null pointer dereferences, out-of-bound accesses, division-by-zero errors in SGX enclave applications \\ \hline
  Jin et al. (2023) \cite{9674230}                        & Fuzzing                                                              & Overflow, underflow, arbitrary value transfer (e.g., reentrancy), suicidal, call injection     \\\hline
  Ha\ss{}ler et al. (2022) \cite{10.1145/3503921.3503924} & Coverage-guided fuzzing                                              & Crash in binary programs                                                                       \\\hline
 Chen et al. (2022) \cite{Chen2022703} & Concolic execution & Fake EOS transfer, fake notification, missing authorization verification, blockinfo dependency, and rollback vulnerabilities \\ \hline
  Zhou et al. (2023) \cite{zhou2023}                      & Bytecode instrumentation, run-time validation, and grey-box fuzzing & Integer and stack overflow \\ \hline
\caption{Vulnerability discovery works that propose a dynamic analysis approach}
\label{tab:vuln-discovery-dynamic}
\end{longtable}
\egroup

%Table~\ref{tab:vulnerability-discovery-sumamry} illustrates the differences in security evaluations performed by various authors. 
A comparative analysis of the mentioned works allows us to observe that there is no standardized approach to comparing the effectiveness of the proposed methods. While some works focus on identifying low-level vulnerabilities such as integer and stack-based overflows, others target vulnerabilities specific to the smart contract domain. 
Some authors use real-world applications for their evaluations, while others employ well-known yet heterogeneous datasets. Evaluating real-world applications effectively demonstrates the capability of security scanners to discover new vulnerabilities. However, this approach does not yield performance metrics such as false positives or negatives, as the exact number of vulnerabilities in the applications under test is unknown. Consequently, each author defines different performance metrics, making it challenging to compare the proposed methods.

This variability in evaluation metrics is a common issue in vulnerability discovery approaches~\cite{10.1145/3474553}, and we encourage future research to address this open problem.

\subsection{Other works}
\label{sec:other}
% comparison and 
Many works focus on aspects of WebAssembly beyond security but are still relevant as they can provide valuable insights for future efforts aimed at enhancing the security of the WebAssembly domain.

\subsubsection{Preliminary works}

Some preliminary studies offer promising results that may serve as a foundation for innovative security improvements.
%POSTER: Leveraging eBPF to enhance sandboxing of WebAssembly runtimes
Abbadini et al. (2023)~\cite{Abbadini20231028} design an approach that combines eBFP features with WebAssembly to enhance host isolation security in existing runtimes.
Narayan et al. (2021)~\cite{narayan2021} provide a tutorial on leveraging WebAssembly's sandboxing feature to run unsafe C code securely.
Stievenart and De Roover (2021)~\cite{wassail2021} introduce ``Wassail'', one of the first static analysis tools for WebAssembly. Their technology enables several static analyses, such as call graphs, control-flow graphs, and dataflow analysis.
Cabrera et al. (2021)~\cite{CabreraArteaga2021} present the first version of a code diversification framework for WebAssembly, which is further extended in a more recent study~\cite{Cabrera-Arteaga2024}. 

% A First Look at Code Obfuscation for WebAssembly
Bhansali et al. (2022)~\cite{10.1145/3507657.3528560} provide a preliminary overview of applying obfuscation techniques to WebAssembly binaries, demonstrating that specific obfuscation approaches can bypass the MINOS cryptojacking detector (Naseem et al. (2021)~\cite{Naseem2021}). 

Zheng et al. (2023)~\cite{Zheng2023867} present a software prototype for dynamic program analysis capable of detecting memory bugs and integer overflows in WebAssembly code, serving as an essential reference for investigating vulnerability discovery in this domain.

\subsubsection{Works about vulnerability discovery approaches}

Many works investigate novel approaches to inspect WebAssembly code and binary structure, laying the groundwork for future vulnerability discovery research.
% taint tracking
William Fu et al. (2018)~\cite{fu2018taintassembly} and Szanto et al. (2018)~\cite{szanto2018taint} provide comprehensive overviews of taint tracking~\cite{Hough2021A} in WebAssembly. Both studies demonstrate promising results in tracking data flows within WebAssembly binaries but do not provide detailed security evaluations. Therefore, further research is needed to assess their effectiveness in real-world scenarios.

% slicing approaches
Stievenart et al. (2022)~\cite{Stievenart20222031} and Stiévenart et al. (2023)~\cite{10336322} introduce methods to minimize WebAssembly binaries through slicing. 
These approaches are useful for optimizing dynamic and dependency analysis of WebAssembly binaries.

% binary variants
Cabrera-Arteaga et al. (2024)~\cite{Cabrera-Arteaga2024} demonstrate how to automatically transform a WebAssembly binary into variants without altering its original functionality. This approach is effective for fuzzing WebAssembly compilers.

Other works can contribute to the realization of static analysis approaches. Notable examples are Wasm Logic, a program logic for first-order WebAssembly proposed by Watt et al. (2019)~\cite{watt2019-2}, and ``Eunomia'' (He et al. (2023) \cite{10.1145/3597926.3598064}), a symbolic execution engine for WebAssembly. 

Stievenart et al. (2020)~\cite{Stievenart202013} present a novel procedure based on static analysis of WebAssembly programs based on compositional information flow. This method enables an efficient and precise summarization of function behaviors, facilitating the detection of information flow violations and potential security breaches. 
%Also, Li and Zhang (2022) \cite{Li2022} introduce a dataflow analysis method that generates Static Single Assignment (SSA) \cite{10.1145/765568.765573} from intermediate representation (IR). 

% Dynamic approaches
Lehmann et al. (2019)~\cite{Lehmann20191045} contribute to the dynamic analysis of WebAssembly with Wasabi, a framework designed to dynamically analyze WebAssembly programs to collect runtime information, enabling the detection of dynamic behaviors such as memory accesses and control flow. 

% Vulnerability discovery and use case
An interesting WebAssembly use case is proposed by Kotenko et al. (2023)~\cite{Kotenko2023}, who devise practical guidelines for implementing secure applications in the power energy sector. Even if WebAssembly is merely mentioned, this work can pave the ground for future research.

Namjoshi et al. (2021)~\cite{10.1007/978-3-030-67067-2_7} define an interesting compiler extension that checks for program correctness against an independent proof validator. Although the proposed contribution is related to revealing bugs that occurred in the optimization process of compiled programs, it could be easily broadened in order to also include security verifications.

To identify bugs in WebAssembly runtimes, Zhou et al. (2023)~\cite{10298359} propose WADIFF, a differential testing framework that could be specialized to find security bugs. 
Debugging code remains an effective approach to manually discover vulnerabilities~\cite{debughafiz}. Lauwaerts et al. (2022)~\cite{10.1145/3546918.3546920} extend the WARDuino WebAssembly microcontroller virtual machine~\cite{LAUWAERTS2024101268} to enable debugging features. Future works may extend this by proposing security methodologies to find vulnerabilities in embedded systems designed with WebAssembly through debugging approaches.

\subsubsection{Comparison with other works}

Another category of works compares WebAssembly security with other technologies. 
Dejaeghere et al. (2023)~\cite{Dejaeghere202335} compare the security features of WebAssembly with those provided by eBPF, a Linux subsystem that allows the safe execution of untrusted user-defined extensions inside the kernel~\cite{Gershuni2019Simple}. 
They demonstrate that different threat models can be defined for these two technologies and emphasize that WebAssembly's design focuses more on security than performance. While WebAssembly permits the execution of unsafe code, unlike eBPF, it reduces security risks and exploitation opportunities by limiting the number of implemented helper system libraries. 

In their study, Pham et al. (2023)~\cite{Pham2023519} assess several non-functional requirements, including performance, maintainability, and security. They demonstrate that WebAssembly is a viable alternative to Docker containers in IoT applications. 
However, the security evaluation is not exhaustive, suggesting the need for further studies to analyze the security disparities between these technologies.

Bastys et al. (2022)~\cite{10.1007/978-3-031-22308-2_5} present a significant contribution with their  SecWasm framework, designed for general-purpose information-flow control in WebAssembly. SecWasm can be instrumented to perform several relevant security controls against WebAssembly programs. While the paper provides a comprehensive description of the approach, it lacks evaluations. Future research could explore and analyze this promising approach further.

In the realm of kernel isolation, Peng et al. (2023)~\cite{10179284} introduce a novel kernel context isolation system that offers enhanced isolation functionality and features an efficient ``implicit'' context switch to minimize performance overhead. Although the paper compares the system to WebAssembly to demonstrate isolation benefits, further evaluation could enhance understanding.

\section{Discussion}
\label{sec:discussion}
In this section, we discuss the results of our survey and summarize the characteristics of the analyzed works. 

%\subsection{Statistics about reviewed works}

Our research reviewed~\totalworks~works. Of these, \categorizedworks~(85,85\%) have been classified into seven categories. The remaining \otherworks~(14,15\%) works do not make significant contributions to the security domain in WebAssembly, yet represent relevant references for novel pathways in related research topics. 

\begin{figure}[h!t]
    \centering
    \includegraphics[width=.8\textwidth]{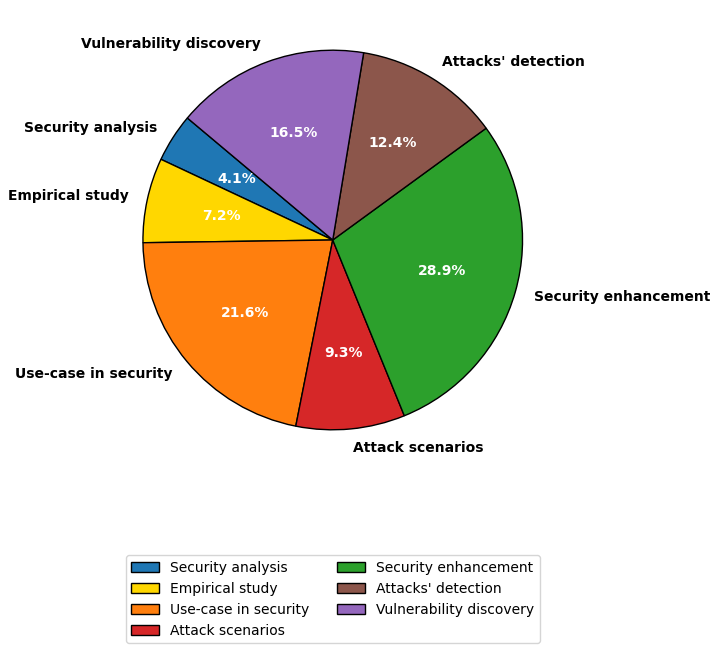}    
    \caption{WebAssembly security research papers categorization}
    \label{fig:pie-wasm}
    \end{figure}

Figure~\ref{fig:pie-wasm} shows the percentage of works for each category. As it is possible to observe, most contributions (28,9\%) aim to harden WebAssembly security. Another large set of works  (21,6\%) leverages the WebAssembly capabilities for implementing security-based use cases. The other relevant areas are related to vulnerability discovery (16,5\%) and attack detection (12,4\%), respectively. 

%\ref{sec:reviewed:appendix} provides a comprehensive list of reviewed works. 

\subsection{WebAssembly security}

Studies that explore the security of WebAssembly through analyses, empirical studies, or by showing attack scenarios led to the following primary conclusions:
\begin{itemize}
    \item Although WebAssembly was designed to provide better isolation and security, the lack of security protection mechanisms in its semantics and compilers has the opposite effect, increasing the risks of memory-based attacks by porting unsafe programs developed with low-level languages such as C into WebAssembly; 

    \item This problem is amplified by the presence of a large number of WebAssembly binaries derived from memory-unsafe languages in real-world applications;

    \item WebAssembly should provide isolation and protect sensitive data from disclosure, but several studies confirm that it is possible to extract sensitive data through specific techniques, such as side-channel attacks based on cache analysis or port contention;

    \item A significant number of WebAssembly binaries in the wild are obfuscated, underscoring the need to expand research into reverse engineering and WebAssembly decompilation; 

    \item The obfuscation applied to WebAssembly has led to a performance reduction in malware detection systems, highlighting the importance of exploring the research field of WebAssembly binary reverse engineering; 

    \item The relevance of crypto-mining activities based on WebAssembly in the world is a contested matter: while some studies underline that it is the primary usage of such technology, others state that less than 1\% of the analyzed binaries are related to crypto-mining activities.
\end{itemize}

\subsection{Detect attacks and discover vulnerabilities in WebAssembly}

Detection efforts primarily focus on cryptojacking attacks, while vulnerability discovery approaches mainly address vulnerabilities in smart contracts. Consequently, there is significant interest in analyzing WebAssembly applications in smart contract environments. As stated in Section~\ref{sec:wasm-usecases-non-security}, there are relevant use-cases in web applications, data visualization, and gaming. Future works should analyze the security impact of WebAssembly in these other domains and propose approaches for addressing security problems. 

Another significant issue is the heterogeneity of used benchmarks and performance metrics. Some authors merely analyze their accuracy, and no common dataset is used for trials and experiments.

The Alexa top one million website benchmark~\cite{alexa} was extensively employed as ``ground truth'' for representing real-world WebAssembly cases. Unfortunately, such a source was retired in 2022 and is unusable for further studies. We encourage researchers to expand their scope by including additional sources to create a more reliable ground truth benchmark.

\subsection{The interest of WebAssembly for cybersecurity use-cases}

\begin{figure}[ht!]
    \centering
    \includegraphics[width=.8\textwidth]{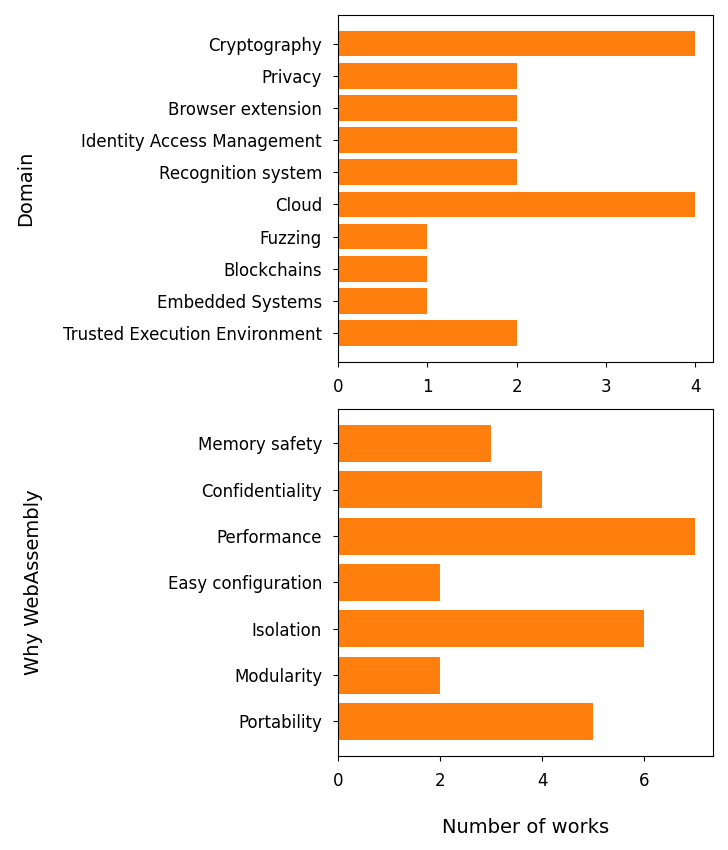}
    \caption{WebAssembly use-cases}
    \label{fig:usecases-chart.png}
    \end{figure}

Figure~\ref{fig:usecases-chart.png} shows the number of works published for each security-related domain and the reasons behind the WebAssembly adoption. 
As observed, WebAssembly is utilized across various application types, with particular relevance in cloud and cryptography domains.
WebAssembly is chosen for several reasons, notably performance, isolation, and portability. Some studies also highlight its adoption to ensure memory safety. However, as previously stated, several studies confirm security gaps in using WebAssembly for both isolation and memory safety, underscoring the necessity for further investigation into the security of these applications.

% PNG IMAGE

\subsection{Enhance WebAssembly security}

\begin{figure}[ht!]
    \centering
    \includegraphics[width=.7\textwidth]{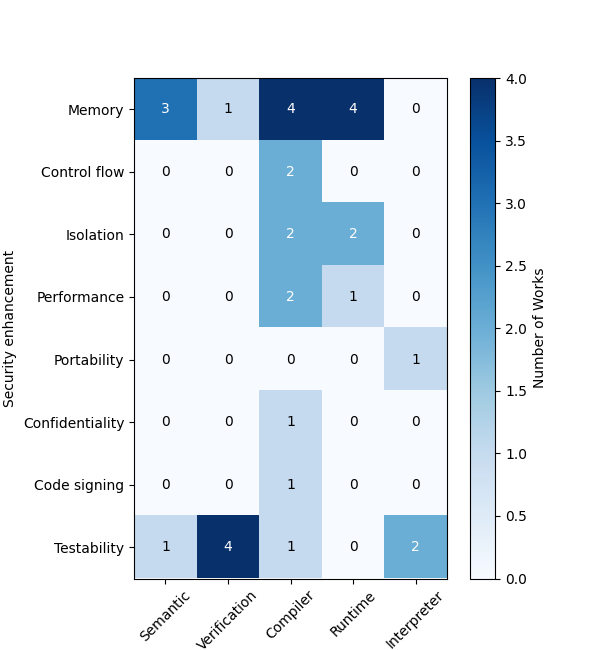}
    \caption{Heatmap for security enhancements works}
    \label{fig:enhance}
    \end{figure}

Figure~\ref{fig:enhance} reports a heatmap illustrating the number of works extending specific components of WebAssembly to enhance their security properties. 

%The picture clearly indicates that making memory management more resilient to attacks, thanks to improvements at the level of both compilers and runtime support, is the focus of the majority of reviewed works. Similar considerations apply to enhancing testability through advanced verification techniques.
What can be inferred from the heat map is that most of the works focus on memory operations and involve modifications to the compiler and runtime. As regards testability, it is analyzed through formal verification methods. Indeed, many improvements could be made through the interpreter since it has less impact compared to changes in the runtime and compiler, thus providing a potential direction for future works. Additionally, confidentiality and code signing are other elements that need to be further explored.

\subsection{Results availability}

For each security category, we analyze whether the authors of relevant works have made available the source code used for the experiments they carried out. The results of this analysis are condensed in Table~\ref{tab:public-available-works}. A detailed list of the publicly available repositories we were able to check is instead reported in Table~\ref{tab:public-work-table} in the Appendix. %~\ref{sec:open-source:appendix}. 
As it is possible to observe, a significant number of works have publicly released their code. This open-source release of data and code will enable researchers to expand the literature on WebAssembly more rapidly and develop novel approaches in the field. We hope that the collection provided in this work will further boost research advancements.  
\begin{table}[ht]
\centering
\begin{tabular}{|M{3,5cm}|M{2cm}|M{2,5cm}|}
\hline
\textbf{Category} &  \textbf{Relevant Works} & \textbf{Public repositories} \\ \hline
Security Analysis   &  \secanalysisworks    & 2\\ \hline
Empirical studies &  \empiricalworks    & 4 \\ \hline
Use case in cybersecurity &  \usecaseworks     & 6 \\ \hline
Attack Scenarios   &  \attackscenarioworks   & 3 \\ \hline
Attacks' detection   &  \attackdetectionworks    & 3 \\ \hline
Security Enhancements  &  \enhancementworks & 17 \\ \hline
Vulnerability Discovery   &   \vulndiscworks    & 9 \\ \hline

\end{tabular}
\caption{Security categories and publicly available works}
\label{tab:public-available-works}
\end{table}

\subsection{Research gaps in WebAssembly and security}

Rethinking the aforementioned considerations, we can summarize the following research gaps: 
\begin{itemize}
    \item WebAssembly has been thoroughly analyzed in the real world, especially through the Alexa \cite{alexa} benchmark, which was retired in 2022. Future studies should investigate additional sources to confirm the empirical results of published works;

    \item Empirical studies do not address the adoption of WebAssembly on the dark web. Given that this technology is used for malicious activities such as evading malware detection and crypto-mining, exploring the relationship between WebAssembly and the dark web could be a compelling research direction.

    \item Crypto-mining and smart contracts are highly relevant in WebAssembly research, with different approaches proposed for detecting attacks and discovering vulnerabilities in these fields. However, empirical studies indicate that WebAssembly is also applied to other types of applications. More studies should investigate this relevance and potentially promote research in unexplored fields where WebAssembly is prominently employed.

    \item WebAssembly is adopted for various security use cases for reasons such as performance, portability, and isolation. However, other interesting applications could be examined. One such application is related to cyber ranges, i.e., realistic offensive and defensive scenarios where people in the security field can be trained~\cite{Yamin2020Cyber}.

    \item  Almost all security analysis works assess the security of C programs compiled for WebAssembly. However, WebAssembly developers write programs in other languages, such as JavaScript and Python. Additionally, the Rust language is now a memory-safe alternative to C. We suggest expanding security research to explore these languages. 

\end{itemize}

\section{Conclusions}
\label{sec:conclusions}

In this work, we provide a comprehensive overview of the current state of security research in WebAssembly, highlighting key studies that address various security aspects of this emerging technology.  We conducted a thorough review of the most recent literature, analyzing a total of~\totalworks~studies, categorizing them into eight distinct categories, and critically evaluating the results. 

Our analysis reveals significant research gaps in the WebAssembly security field, which we believe future studies should aim to address. Specifically, we identify the lack of exploration of cyber-ranges as a promising yet underutilized security use case in WebAssembly, which has the potential to revolutionize the way we approach security testing and training. To contribute to this area, we plan to investigate the benefits of using WebAssembly in cyber-ranges by comparing our previous research~\cite{8169747, Caturano2020, Luise202180} with a cyber-range solution developed entirely using this technology. 

We also highlight the potential for WebAssembly to enable more efficient and effective security testing and discuss the importance of considering the security implications of WebAssembly's unique features, such as its sandboxed execution environment and memory safety guarantees. 

We hope that this work will be useful to researchers seeking to make a meaningful impact in the challenging field of enhancing the security features of WebAssembly and provide valuable insights for developers and practitioners looking to leverage WebAssembly for their security applications.

 \bibliographystyle{unsrtnat} 
 \bibliography{bib}





% \end{thebibliography}

\newpage
\section*{Appendix: publicly available source code repositories}
\label{sec:open-source:appendix}
% LONGTABLE
\bgroup
\centering
\begin{longtable}{|M{0.2\linewidth}|M{0.7\linewidth}|}
\hline

\multicolumn{2}{|c|}{\textbf{Security analysis}} \\ \hline
\emph{Work}  & \emph{Publicly available source} \\ \hline
Lehmann et al. (2020) \cite{Lehmann2020217} & \url{https://github.com/sola-st/wasm-binary-security} \\ \hline
Stiévenart et al. (2022) \cite{Stiévenart20221713} & \url{https://figshare.com/articles/dataset/SAC_2022_Dataset/17297477} \\ \hline

\multicolumn{2}{|c|}{\textbf{Empirical studies}} \\ \hline
\emph{Work}  & \emph{Publicly available source} \\ \hline
Romano et al. (2021) \cite{9678776} & \url{https://github.com/sola-st/WasmBench} \\ \hline
Hilbig et al. (2021) \cite{10.1145/3442381.3450138} & \url{https://wasm-compiler-bugs.github.io/} \\ \hline
Konoth et al. (2018) \cite{10.1145/3243734.3243858} & \url{https://github.com/vusec/minesweeper} \\ \hline

Tsoupidi et al. (2021) \cite{tsoupidi} & \url{https://github.com/romits800/Vivienne} \\ \hline

\multicolumn{2}{|c|}{\textbf{Use case in cybersecurity}} \\ \hline
\emph{Work}  & \emph{Publicly available source} \\ \hline

Narayan et al. (2020) \cite{Narayan2020699} & \url{https://github.com/PLSysSec/rlbox-usenix2020-aec} \\ \hline
Izquierdo Riera et al. (2023) \cite{icissp23} & \url{https://github.com/clipaha/clipaha} \\ \hline
Goltzsche et al. (2019) \cite{10.1145/3361525.3361541} & \url{https://github.com/ibr-ds/AccTEE} \\ \hline
M\'{e}n\'{e}trey et al. (2024) \cite{menetrey2023} & \url{https://github.com/JamesMenetrey/unine-opodis2023} \\ \hline
Attrapadung et al. (2018) \cite{Attrapadung2018} & \url{https://github.com/herumi/mcl} \\ \hline
Baumgärtner et al. (2019) \cite{Baumgärtner2019} & \url{https://github.com/stg-tud/bp7eval} \\ \hline 

\multicolumn{2}{|c|}{\textbf{Attack scenarios}} \\ \hline
\emph{Work}  & \emph{Publicly available source} \\ \hline
Cabrera-Arteaga et al. (2023) \cite{CABRERAARTEAGA2023103296} & \url{https://github.com/ASSERT-KTH/wasm_evasion} \\ \hline
Easdon et al. (2022) \cite{277134} 
 & \url{https://github.com/libtea/frameworks} \\ \hline
 Oz et al. (2023) \cite{291205} & \url{https://github.com/cslfiu/RoB_Ransomware_over_Modern_Web_Browsers} \\ \hline

\multicolumn{2}{|c|}{\textbf{Attacks' detection}} \\ \hline
\emph{Work}  & \emph{Publicly available source} \\ \hline

Kharraz et al. (2019) \cite{10.1145/3308558.3313665} & \url{https://github.com/teamnsrg/outguard} \\ \hline
Bian et al. (2020)~\cite{Bian20203112}              &  \url{https://github.com/cuhk-seclab/MineThrottle} \\  \hline
Romano et al. (2020) \cite{9286112} & \url{https://miner-ray.github.io} \\ \hline
 
\multicolumn{2}{|c|}{\textbf{Security enhancements}} \\ \hline
\emph{Work}  & \emph{Publicly available source} \\ \hline

Peach et al. (2020) \cite{Peach20203492} & \url{https://github.com/gwsystems/awsm/} \\ \hline
Zhang et al. (2021) \cite{Zhang2021} & \url{https://github.com/Zhiyi-Zhang/PS-Signature-and-EL-PASSO} \\ \hline
Vassena et al. (2021) \cite{10.1145/3434330} & \url{https://github.com/PLSysSec/blade} \\ \hline
Narayan et al. (2021a) \cite{272260} & \url{https://github.com/PLSysSec/swivel} \\ \hline
Lei et al. (2023) \cite{Lei2023904} & \url{https://github.com/PKU-ASAL/PKUWA} \\ \hline
Watt et al. (2023) \cite{10.1145/3591224} & \url{https://github.com/WasmCert/WasmCert-Isabelle} \\ \hline
Menetrey et al. (2021) \cite{Menetrey2021205} & \url{https://github.com/jamesmenetrey/unine-twine} \\ \hline
Geller et al. (2024) \cite{10.1145/3632922} & \url{https://dl.acm.org/do/10.1145/3580426/full/} \\ \hline
Johnson et al. (2023) \cite{10179357} & \url{https://github.com/PLSysSec/wave} \\ \hline
Narayan et al. (2023) \cite{Narayan2023266} & \url{https://github.com/PLSysSec/hfi-root} \\ \hline
Watt et al. (2019) \cite{Watt2019} & \url{https://github.com/PLSysSec/ct-wasm} \\ \hline
Romano et al. (2022) \cite{9833626} & \url{https://github.com/js2wasm-obfuscator/translator} \\ \hline
Jay Bosamiya et al. (2022) \cite{279990} & \url{https://github.com/secure-foundations/provably-safe-sandboxing-wasm-usenix22} \\ \hline 
Michael et al. (2023) \cite{Michael2023425} & \url{https://github.com/PLSysSec/ms-wasm} \\ \hline
Watt et al. (2021) \cite{watt2021} & \url{https://github.com/WasmCert} \\ \hline
Menetrey et al. (2022) \cite{Menetrey20221177} & \url{https://github.com/JamesMenetrey/unine-watz} \\ \hline
Schrammel et al. (2020) \cite{255298} & \url{https://github.com/IAIK/Donky} \\ \hline

\multicolumn{2}{|c|}{\textbf{Vulnerability discovery}} \\ \hline
\emph{Work}  & \emph{Publicly available source} \\ \hline
Khan et al. (2023) \cite{10190488} & \url{https://github.com/purseclab/FuzzSGX} \\ \hline
Li et al. (2022) \cite{Li2022746} & \url{https://github.com/lwy0518/datasets_results} \footnote{The work only provides dataset and results, but no source code} \\ \hline

Jiang et al. (2021) \cite{Jiang2021926} & \url{https://github.com/gongbell/WANA} \\ \hline
Brito et al. (2022) \cite{Brito2022} & \url{https://github.com/wasmati/wasmati} \\ \hline
Lehmann et al. (2021) \cite{lehmann2021fuzzm} & \url{https://github.com/fuzzm/fuzzm-project} \\ \hline 
Haßler and Maier (2022) \cite{10.1145/3503921.3503924} &  \url{https://github.com/fgsect/WAFL} \\ \hline
Quan et al. (2019) \cite{quan2019evulhunter} & \url{https://github.com/EVulHunter/EVulHunter} \\ \hline
Chen et al. (2022) \cite{Chen2022703} & \url{https://github.com/wasai-project/wasai} \\ \hline
Zhou and Chen (2023) \cite{zhou2023} & \url{https://github.com/HuskiesUESTC/AntFuzzer-WASMOD} \\ \hline

\multicolumn{2}{|c|}{\textbf{Other works}} \\ \hline
\emph{Work}  & \emph{Publicly available source} \\ \hline
William Fu et al. (2018) \cite{fu2018taintassembly} & \url{https://github.com/wfus/WebAssembly-Taint} \\ \hline
Namjoshi et al. (2021) \cite{10.1007/978-3-030-67067-2_7} & \url{https://github.com/nokia/web-assembly-self-certifying-compilation-framework} \\ \hline

Narayan et al. (2021) \cite{narayan2021} & \url{https://rlbox.dev/} \\ \hline 
Katherine Hough and Jonathan Bell (2021) \cite{Hough2021A} & \url{https://doi.org/10.6084/m9.figshare.16611424.v1} \\ \hline
Stievenart et al. (2021,2022) \cite{wassail2021, Stievenart20222031} & \url{https://github.com/acieroid/wassail}  \\ \hline 

Stiévenart et al. (2023) \cite{10336322}  & \url{https://doi.org/10.5281/zenodo.8157269} \\ \hline 

Cabrera Arteaga et al. (2021) \cite{CabreraArteaga2021} & \url{https://github.com/ASSERT-KTH/slumps/tree/master/crow} \\ \hline
He et al. (2023) \cite{10.1145/3597926.3598064} & \url{https://github.com/PKU-ASAL/Eunomia-ISSTA23} \\ \hline

Stievenart et al. (2020) \cite{Stievenart202013} & \url{https://github.com/acieroid/wassail/releases/tag/scam2020} \\ \hline

Lehmann et al. (2019) \cite{Lehmann20191045} & \url{http://wasabi.software-lab.org/} \\ \hline
Bastys et al. (2022) \cite{10.1007/978-3-031-22308-2_5} & \url{https://github.com/womeier/secwasm} \\ \hline

Zhou et al. (2023) \cite{10298359} & \url{https://github.com/erxiaozhou/WaDiff} \\ \hline

Cao et al. (2023) \cite{10.1007/978-3-031-44245-2_8} & \url{https://github.com/security-pride/BREWasm} \\ \hline

Dejaeghere et al. (2023) \cite{Dejaeghere202335}  & \url{https://gist.github.com/Rekindle2023/ca6a072205698925fa80f928eebe172e} \\ \hline

Peng et al. (2023) \cite{10179284} & \url{https://github.com/rssys/uswitch} \\ \hline
Cabrera-Arteaga et al. (2024) \cite{Cabrera-Arteaga2024} & \url{https://github.com/bytecodealliance/wasm-tools/tree/main/crates/wasm-mutate} \\ \hline

\caption{Publicly available source codes of analyzed works All the provided links were valid in May 2024. Other works are not included in the analysis}
\label{tab:public-work-table}
\end{longtable}
\egroup

\end{document}